\immediate\write18{makeindex \jobname.nlo -s nomencl.ist -o \jobname.nls} 
\documentclass[preprint,1p,sort&compress,times]{elsarticle}

\usepackage{amssymb}
\usepackage{amsmath}
\usepackage{makecell}
\usepackage{booktabs}
\usepackage{tabularx}
\usepackage{multirow}
\usepackage{siunitx} 

\usepackage{lineno}

\usepackage{framed} 
\usepackage{multicol} 
\usepackage{nomencl} 
\makenomenclature
\renewcommand*\nompreamble{\begin{multicols}{2}}
\renewcommand*\nompostamble{\end{multicols}}
\usepackage{etoolbox} 
\renewcommand\nomgroup[1]{%
  \item[\bfseries
  \ifstrequal{#1}{A}{Abbreviations}{%
  \ifstrequal{#1}{S}{Sets and indices}{%
  \ifstrequal{#1}{P}{Parameters}{%
  \ifstrequal{#1}{D}{Decision Variables}{%
  \ifstrequal{#1}{V}{Other Variables}{}}}}}%
]}%

\journal{Applied Energy}

\begin{document}

\begin{frontmatter}



\title{Optimal wind farm energy and reserve scheduling incorporating wake interactions}


\author[1,2]{Marin Mabboux-Fort\corref{cor1}} 
\ead{marin.mabboux-fort@sig-ge.ch}

\author[2]{Majid Bastankhah} 
\ead{majid.bastankhah@durham.ac.uk}

\author[2]{Peter C Matthews} 
\ead{p.c.matthews@durham.ac.uk}

\author[3]{Mokhtar Bozorg} 
\ead{mokhtar.bozorg@heig-vd.ch}

\cortext[cor1]{Corresponding author}

\affiliation[1]{organization={Services Industriels de Genève},
            addressline={Chemin du Château-Bloch 2}, 
            city={Le Lignon},
            postcode={1219}, 
            state={Geneva},
            country={Switzerland}}
            
\affiliation[2]{organization={Department of Engineering, University of Durham },
            addressline={Stockton Road}, 
            city={Durham},
            postcode={DH1 3LE}, 
            country={UK}}

\affiliation[3]{organization={School of Engineering and Management Vaud (HEIG-VD), University of Applied Sciences and Arts Western Switzerland (HES-SO)},
            addressline={Route de Cheseaux 1}, 
            city={Yverdon-les-Bains},
            postcode={1401}, 
            state={Vaud},
            country={Switzerland}}

\begin{abstract}
This paper proposes a novel approach for optimal energy and reserve scheduling of wind farms by explicitly modelling wake interactions to enhance market participation and operational efficiency.
Conventional methods often neglect wake effects, relying on power curve estimations that represent an upper limit and reduce market performance.
To address this, a two-stage stochastic programming framework is developed, integrating a wake-aware power estimation model within the FLORIS simulation software.
Wind and reserve uncertainties are addressed through scenario generation and reduction, enabling wind power producers to optimise participation in day-ahead energy and ancillary services markets, with particular focus on the Frequency Restoration Reserve (FRR).
The wake-aware model provides more realistic power output predictions based on site-specific wind and atmospheric conditions, improving scheduling accuracy and reducing imbalance penalties. 
Wake steering is further employed to mitigate wake-induced losses and increase income through participation in ancillary services. 
The proposed approach is evaluated through a case study of the London Array offshore wind farm participating in the Great Britain (GB) electricity markets.
Results show that conventional methods estimate production 12-13\% higher, leading to imbalance penalties and 3\% lower revenue compared with the wake-aware approach accounting for wake interactions. 
Moreover, the steering-enhanced approach yields an additional 1-2\% increase in income relative to the wake-aware baseline.
These findings underscore the value of accounting for wake interactions in wind farm scheduling and demonstrate the economic and operational benefits of active wake management, offering insights for improving grid stability and profitability as wind penetration continues to rise.
\end{abstract}

\begin{graphicalabstract}
\end{graphicalabstract}

\begin{highlights}
\item Wind farm scheduling model incorporates wake effects and wake steering control.
\item Two-stage stochastic framework with wake-aware power estimation using FLORIS. 
\item Traditional wind power estimation methods can overestimate production by 12-13\%.
\item Wake steering improves wind farm income by 1-2\% compared to non-steering methods.
\end{highlights}

\begin{keyword} 
wind power \sep electricity markets \sep ancillary services \sep stochastic optimisation \sep bidding strategy \sep wake effects
\end{keyword}

\end{frontmatter}



\nomenclature[A]{DA}{Day-ahead}
\nomenclature[A]{FCR}{Frequency Containment Reserve}
\nomenclature[A]{FR}{Fast Reserve}
\nomenclature[A]{FRR}{Frequency Restoration Reserve}
\nomenclature[A]{GB}{Great Britain}
\nomenclature[A]{MFR}{Mandatory Frequency Response}
\nomenclature[A]{NESO}{National Energy System Operator}
\nomenclature[A]{TSO}{Transmission System Operator}
\nomenclature[A]{TI}{Turbulence Intensity}
\nomenclature[A]{WF}{Wind Farm}
\nomenclature[A]{WPP}{Wind Power Producer}
\nomenclature[A]{WT}{Wind Turbine}

\nomenclature[S]{\(s/S\)}{Scenario index/ set of scenarios}
\nomenclature[S]{\(w_s\)}{Wind speed signal}
\nomenclature[S]{\(w_d\)}{Wind direction signal}
\nomenclature[S]{\(t\)}{Time index (hour)}
\nomenclature[S]{\(\phi\)}{Set of decision variables}

\nomenclature[P]{\(E_{gen}\)}{Generated Energy frm the WF}
\nomenclature[P]{\(\lambda_e(t)\)}{Day-ahead energy market price during time interval $t$ [\pounds/MWh]}
\nomenclature[P]{\(\lambda_{mfr}^\uparrow(t)\)}{Holding payment of upward MFR during time interval $t$ [\pounds/MW/h]}
\nomenclature[P]{\(\lambda_{fr,a}^\uparrow(t)\)}{Availablity fee of FR during time interval $t$ [\pounds/MW/h]}
\nomenclature[P]{\(\lambda_{fr,u}^\uparrow(t)\)}{Utilisation payment of FR during time interval $t$ [\pounds/MWh]}
\nomenclature[P]{\(\lambda_{b,e}(t)\)}{Energy imbalance settlement price during time interval $t$ [\pounds/MWh]}
\nomenclature[P]{\(\lambda_{b,fr}(t)\)}{FR imbalance settlement price during time interval $t$ [\pounds/MWh]}
\nomenclature[P]{\(\rho^s\)}{Probability of occurence of scenario $s$}
\nomenclature[P]{\(\Delta t_{fr}^s(t)\)}{FR duration during time interval $t$ at scenario $s$ [h]}
\nomenclature[P]{\(P_{max}(t)\)}{Maximum forecasted power of  the WF's generation [MW]}
\nomenclature[P]{\(P_{max}^s(t)\)}{Total available output power of  the WF's generation at scenario $s$ [MW]}
\nomenclature[P]{\(P_{min}(t)\)}{Minimum limit of  the WF's generation [MW]}
\nomenclature[P]{\(P_{fr}^{min\uparrow}\)}{Minimum FR power requirement for participation in the FR market [MW]}

\nomenclature[D]{\(P_e(t)\)}{Power bid regarding the day-ahead energy market at time $t$ [MW]}
\nomenclature[D]{\(P_{mfr}^\uparrow(t)\)}{Power bid regarding the day-ahead upward MFR market at time $t$ [MW]}
\nomenclature[D]{\(P_{fr}^\uparrow(t)\)}{Power bid regarding the day-ahead FR market at time $t$ [MW]}
\nomenclature[D]{\(\Delta P_{e}^s(t)\)}{Energy redispatch at time $t$ of scenario $s$ [MW]}
\nomenclature[D]{\(\Delta P_{fr}^s(t)\)}{FR power redispatch at time $t$ of scenario $s$ [MW]}

\nomenclature[V]{\(P_{gen}\)}{Generated WF power}
\nomenclature[V]{\(\prod\)}{Total revenue of the WPP in the energy and reserve market}
\nomenclature[V]{\(\prod_h\)}{Total expected hourly income of the WPP after balancing stage}
\nomenclature[V]{\(\prod_d\)}{Total expected daily income of the WPP}

\begin{table*}[!t]
\begin{framed}
\fontsize{8}{9}\selectfont
\printnomenclature
\end{framed}
\end{table*}

\section{Introduction} 
\label{intro}
Growth of energy consumption worldwide has led to an extensive use of fossil fuels from natural exhaustive resources \cite{leao_future_2011}.
The finite nature of these resources and environmental concerns have led different countries and regions to introduce policies aimed at reducing the environmental footprint of the energy sector and increasing renewable energy penetration.
The EU has committed to become the global leader in decarbonising the power system, and wind power is essential in reaching its carbon-neutral target by 2050 \cite{muneer_wind_2022}.
Today, wind power generation supplies more than 10\% of European consumption, and is expected to grow to 33\% in 2030 \cite{noauthor_windeurope_2016}.
Furthermore, the installed global wind power capacity is increasing rapidly, reaching 906 GW by the end of 2022 \cite{noauthor_global_2023}, and rising investments worldwide in renewable energy projects indicates an increase of their penetration in power systems \cite{abbasi_coordinated_2019}.
However, effective integration of wind energy into the power system can raise concerns about grid stability and reliability \cite{yuan_race_2022}.
Despite its notable economic and environmental advantage, the inherent intermittency of wind adversely affects the stability and reliability of power systems to a higher degree \cite{gao_different_2019}.
Transmission System Operators (TSOs) require a range of system services to ensure the stable and reliable operation of the electrical power system, and it remains unclear how system stability can be maintained with high penetration of non-synchronous renewable energy sources \cite{banshwar_renewable_2017}.
As the share of uncertain generation rises, the need for more responsive and costly system services correspondingly increases \cite{tsimopoulos_optimal_2019}. 
This is already a challenge in regions that have a significant portion of wind energy, such as Great Britain (GB), Spain, Ireland, and Denmark \cite{world_wind_energy_association_world_2011}.
High wind penetration levels in the aforementioned countries have led TSOs to require wind farms (WFs) to actively balance the grid through market participation and provision of ancillary services, roles traditionally fulfilled by conventional power plants \cite{li_adaptive_2017}.
Consequently, wind power generation, which has typically been optimised for maximum power output and cost efficiency, must now be adjusted to meet system requirements \cite{siniscalchi-minna_wind_2019}.
These new requirements have forced wind turbine (WT) manufacturers to develop and implement control methodologies to provide ancillary services, and it seems possible that WTs could be more effective at providing some of them than traditional power plants, which presents good opportunities for both TSOs and wind power producers (WPPs) \cite{aho_tutorial_2012}.

\subsection{Wind power control}
\label{wind power control}
The conversion of wind energy into electrical power by a WT generates a downstream wake characterised by reduced velocity and increased turbulence.
Wake effects correspond to reduction in wind speed and the increase in turbulence intensity downstream of operating turbines.
These wake effects influence the airflow within the WF and can affect the performance of nearby turbines. 
In WFs,  wakes can interfere with adjacent turbines, reducing their ability to extract power  and increasing fatigue loads \cite{porte-agel_wind-turbine_2020}.
The conventional approach to WF control operates at the individual turbine level.
Known as greedy control, it employs maximum power point tracking (MPPT), where each turbine operates independently without considering how the turbine performance impacts other adjacent turbines \cite{njiri_state---art_2016}.
MPPT is typically implemented through a control algorithm that optimises power output based on sensor inputs such as wind direction and rotor speed, using  passive stall with fixed blade pitch or active pitch control to adjust blades to stall or feather \cite{burton_wind_2021}. 
However, in practice, the operational conditions of WTs significantly influence the development of their wakes, which in turn affects the energy available to downwind turbines.
The wake can be broadly divided into two regions: the near wake, where flow is dominated by turbine-induced forces and coherent structures, and the far wake, where atmospheric turbulence plays a larger role in promoting mixing and wake recovery. 
Downwind turbines are typically located in the far wake region, making their performance highly dependent on both the upstream turbine operation and the ambient turbulence conditions \cite{stevens_flow_2017}.
The early years of wind energy technology focused primarily on individual turbines, with research on turbine arrays gaining prominence more recently \cite{meyers_wind_2022}.
Wake management aims to maximise power, extend turbine lifetimes, and reduce maintenance, amongst others.
Since wake flows contribute significantly to fatigue loads, effective management can also help redistribute these loads, potentially improving turbine longevity without sacrificing power output.
Advances in computing, sensors, and measurement techniques have facilitated the development of effective wake management strategies. 
As a result, there is an increasing focus on active farm control strategies such as Individual Pitch Control (IPC), Axial Induction Control (AIC), and wake steering. 
Among these,  wake steering has emerged as the most promising wake management technique \cite{houck_review_2022}.
In this strategy, upstream turbines are intentionally yawed to redirect their wakes away from downstream turbines.
While yawing a turbine reduces its own power production, it can enhance the power output of downwind turbines, improving overall farm performance, as demonstrated in several studies (see the review of \cite{houck_review_2022} and references therein).

\subsection{Electricity markets}
\label{electricity markets}
With the restructuring and liberalisation of electricity markets, WPPs can participate in them to maximise their profits \cite{shafie-khah_optimal_2014}.
Wind power can participate in both energy and ancillary services markets, with strong interactions between the two that affect each other's participation \cite{kirschen_fundamentals_2004}.
In energy markets, wind power is typically sold through long-term Power Purchase Agreements (PPAs), notably in the US and Nordic systems \cite{botterud_wind_2010, wiklund_potential_2021}. 
However, some wind power is traded in Day-ahead (DA) and Intraday (ID) markets, where it faces price fluctuations and potential imbalance penalties, necessitating research into optimal bidding strategies \cite{botterud_risk_2010}.
WFs are increasingly required to provide reserve capacity services for frequency control (hereinafter, reserve), which ensures the balance of active power production and consumption to maintain system frequency.
These services are categorised by their response time and duration \cite{canizes_mixed_2013} and include: Frequency Containment Reserve (FCR), or primary control, which stabilises frequency within seconds by releasing active power reserves; Frequency Restoration Reserve (FRR), or secondary control, which restores the frequency to its nominal value within seconds to 15 minutes via automatic (aFRR) or manual (mFRR) activation; and Replacement Reserve (RR), or tertiary control, activated from 15 minutes to hours to support FRR restoration and prepare for future imbalances \cite{holttinen_ancillary_2012}. 
The provision of ancillary services on the spot market, where offers fluctuate based on energy market interactions and varying daily needs, requires the joint optimisation of energy sales and ancillary services. 
Simultaneous and optimal participation in DA energy and ancillary services markets has been shown to enhance WF profitability and mitigate associated risks \cite{shafie-khah_optimal_2014}.

While ancillary markets for renewable energy sources exist in European countries, the need to develop optimal strategies for wind power participation is increasingly critical,  and the challenge remains to effectively incorporate the ability of wind power to provide and accurately schedule ancillary services \cite{kayedpour_optimal_2023}, as high renewable energy penetration in power systems will require significant participation from renewable sources to maintain grid stability and energy security \cite{ebrahimi_provision_2020}.
Wind power participation in GB and Spanish reserve markets was analysed in \cite{edmunds_participation_2019}, and the potential entry of wind power in Swedish markets was explored in \cite{wiklund_potential_2021}.
The technical capacity of WTs to provide ancillary services has been extensively documented \cite{attya_review_2018,cole_critical_2023,altin_optimization_2018,liu_active_2019,kim_optimal_2017,yao_optimized_2019,zhao_fatigue_2017,aemo_wind_2013}, and the potential of using wind power for provision of various ancillary services has been or is being explored in several reports and pilot projects around the world \cite{noauthor_wingrid_2021,nedd_operating_2020,ramboll_ancillary_2019,schipper_recommendations_2019,ela_active_2014,denholm_introduction_2019, faiella_capabilities_2013}.
These studies mainly facilitated the participation of wind power in reserve market using turbine-level control (e.g. pitch and torque control), with less emphasis on farm-level control strategies and their impact on reserve market participation. 

\subsection{Research gap and paper contributions} 
\label{research gap}
Although wind power has been shown to provide ancillary services, wake effects are often overlooked in short-term electricity market participation, leading to inaccuracies in production estimates.
Existing studies typically focus either on optimal market bidding or on WF operation strategies, often neglecting wake effects or market constraints \cite{kayedpour_optimal_2023}.
Studies such as \cite{shafie-khah_optimal_2014} or \cite{kayedpour_optimal_2023} propose optimal operational strategies for WFs, including participation in forward, DA, and balancing markets, as well as wake management for reserve provision.
These studies combine stochastic wind distribution models with manufacturer-reported turbine power curve data to enable short-term forecasting of WF power output.
In this approach, called "Power Curve" in this paper, wind power output is treated as a function of incoming wind speed, based on the turbine power curve and the number of turbines in the farm. 
However, it neglects wake effects and overlooks induced losses, thus representing an upper limit of farm-generated power, which exceeds actual power production due to wake losses.
Other works, such as \cite{botterud_risk_2010}, base wind power production on WF production forecasts.
This approach, referred to as "Baseline" in this paper, accounts for wake losses but does not incorporate wake management techniques as WFs currently lack such strategies, potentially underestimating power production.
Thus, it represents a lower limit of farm-generated power.

The review of existing methodologies for forecasting wind energy production and market participation reveals the limitations of the "Power Curve" approach and emphasises the need to account for wake effects in market participation strategies.
This underscores a research gap between two distinct research streams: electricity market operations and wind power control strategies.
The first stream focuses on maximising profits through participation in energy and ancillary service markets. 
These works typically employ simplified power curve models and often overlook the impact of wake interactions.
Conversely, the second stream concentrates on modelling wake effects and developing active wake management control strategies to optimise WF performance.
However, these studies rarely address the complexities of energy and ancillary service markets, and often do not explore how wake redirection techniques can be integrated into market-based operational strategies to enhance profitability.
Only a limited number of recent works have begun to address this intersection; however, to the best of the authors’ knowledge, a comprehensive formulation that explicitly integrates wake effects, particularly wake steering strategies, within a profit-oriented market participation framework remains limited.
For instance, \cite{starke_yaw-augmented_2023}  introduces a WF control strategy that combines wake steering with pitch control to enhance power reference tracking. 
The study demonstrates that this method improves WF power delivery and regulation capabilities, leading to financial benefits.
By optimising power supply in the energy market while maintaining regulation services, this method enhances revenue potential for WFs providing frequency regulation services.
Building upon previous research, this paper proposes an interdisciplinary approach that bridges the gap between aerodynamic wake modelling and market-oriented decision-making. 
Specifically, it addresses the disconnect between the "Power Curve" and "Baseline" methods by incorporating wake effects into wind power estimation and actively mitigating them through wake management.
Expanding on the work of \cite{starke_yaw-augmented_2023}, this study advances the field by conducting a market analysis to assess the financial potential of wake steering in WF participation in short-term electricity markets. 
Although the formulation is intentionally kept simple to ensure clarity and tractability, it provides a foundational framework for coupling aerodynamic wake modelling with market optimisation and can serve as a basis for more complex future developments.

To support WF decision-making in these markets, a two-stage stochastic programming framework is developed for energy and reserve scheduling in DA markets. 
The contributions of this paper are three-fold:
\begin{enumerate}[1.]
    \item  Wake-aware power estimation: Unlike conventional methods that overlook wake interactions, this study uses state-of-the-art engineering wake models to model wake effects in WF power estimation for market participation.
    \item Wake steering for enhanced efficiency: Wake steering is employed as an active control strategy to mitigate wake-induced losses, optimising both WF efficiency and market performance. 
    \item Extended market analysis: While most studies primarily explored WF participation in the FCR market, this work expands the analysis to the FRR market, addressing wake dynamics in this broader context. 
\end{enumerate}
To better highlight the contribution of this paper, Table \ref{tab:summary} summarises the differences between the "Power Curve", "Baseline", and "Proposed" methodologies.
\begin{table}[t]
\centering
\begin{tabular}{ccccc}
\toprule
\thead[l]{Method\\ {[References]}} & \thead{Wake\\ effects\\ included} & \thead{Wake\\ management\\ applied} & \thead{Power\\ estimation} & \thead{Key features\\ (+/-)} \\
\midrule
\thead[l]{Power Curve\\ \cite{shafie-khah_optimal_2014,kayedpour_optimal_2023,hosseini_advanced_2020,morales_short-term_2010,zhang_optimal_2019,rahimiyan_evaluating_2011,soares_optimal_2016}} & \thead{No} & \thead{No} & \thead{Optimistic\\ (upper limit)} & \thead[l]{+ Simple; wind speed only\\ - May overestimate; penalisation risk} \\ 
\thead[l]{Baseline\\ \cite{botterud_risk_2010,vilim_wind_2014,liang_increased_2011,soares_optimal_2017}} & \thead{Yes} & \thead{No} & \thead{Conservative\\ (lower limit)} &  \thead[l]{+ Simple; no layout data\\ - Potential lost revenue} \\ 
\thead[l]{Proposed} & \thead{Yes} & \thead{Yes} &  \thead{Realistic/\\ Enhanced}&  \thead[l]{+ Optimised output via wake steering\\ - Requires wind \& layout data} \\
\bottomrule
\end{tabular}
\caption{Differences between the proposed method over the approaches presented in the literature regarding forecasting wind energy production and market participation.}\label{tab:summary}
\end{table}

The remainder of this paper is structured as follows. 
Section \ref{methods} outlines the proposed methodology.
The stochastic scenario generation and reduction process is described in Section \ref{scenarios}.
In Section \ref{optimisation}, the optimisation problem is formulated, defining the objective function and constraints.
Section \ref{results and discussion} presents the case study and corresponding numerical results to assess the effectiveness of the proposed method, highlighting its primary advantages over benchmark approaches as well as its limitations. 
Finally, Section \ref{conclusion} summarises the key findings and concludes the paper. 

\section{Methodology overview}
\label{methods}
Recent developments in electricity market regulations, coupled with advancements in WF control technologies, have motivated WPPs to engage jointly in DA energy and reserve markets.
In this context, several studies have investigated optimal bidding strategies for WPPs in these markets (see \cite{hosseini_advanced_2020} and references therein).
This paper adopts a two-stage stochastic programming framework to account for wind and reserve uncertainties, which are modelled as time-variant scenarios.
The use of discrete scenarios derived from available data to simulate uncertain parameters is a common approach in power system applications \cite{abedinia_optimal_2019, khodaei_fuzzy-based_2018}.
In the first stage, the objective is to determine the optimal allocation of power across the different markets to maximise the WPP’s expected profit, based on forecasted wind generation and reserve requirements.
Bidding decisions for energy production and reserve capacity are scheduled a day before activation and cannot be adjusted afterward.
Therefore, the second stage accounts for uncertainty by evaluating a set of potential scenarios and their expected values.
This allows for a more robust decision-making process that mitigates the effects of variability on market participation,  increasing the likelihood of fulfilling contracted power commitments in real-time.

\begin{figure}[t] 
\centering
\includegraphics[height=0.6\textheight,keepaspectratio]{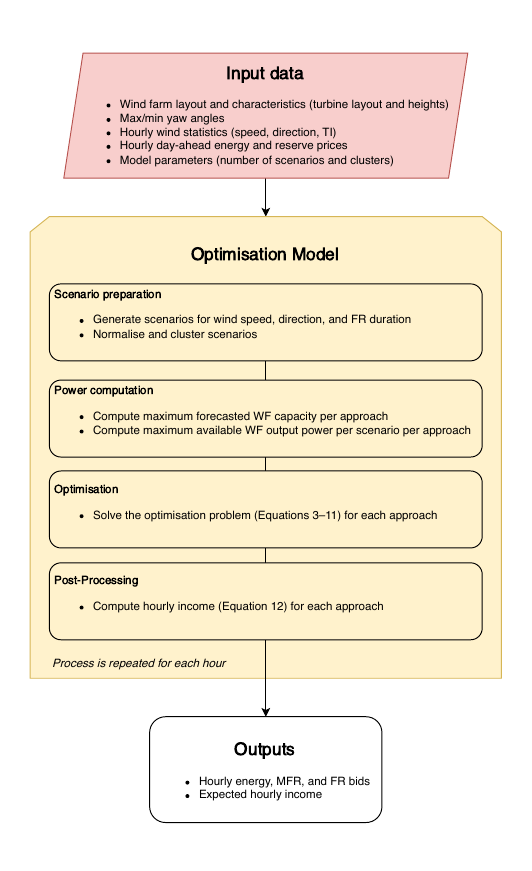}
\caption{Proposed optimal energy and reserve scheduling strategy}\label{fig:flowchart}
\end{figure}

Scenario generation for wind speed and direction enables WF power estimation while accounting for wake effects.
To evaluate the impact of wakes on WF power output, a wake model is required to characterise the velocity deficit and turbulence levels.
For WF optimisation tasks, high-fidelity numerical simulations are often prohibitively expensive. 
As a result, a common approach is to use engineering wake models, which simplify the governing flow equations to efficiently predict both the velocity deficit and wake deflection caused by yaw misalignment in wake steering strategies. 
The wake model employed in this study is the Cumulative Curl model \cite{bay_addressing_2023,bastankhah_analytical_2021}, which is the default model in FLORIS at the time of writing.
FLORIS is a Python-based WF simulation software library designed to evaluate turbine-turbine interactions using steady-state engineering wake models \cite{noauthor_floris_2023}. 
It features an optimisation routine that determines the optimal yaw angle distribution for power maximisation based on wind speed, wind direction, and turbulence intensity. 
This integration allows for a more accurate representation of WF power output. 
The proposed operational strategy is shown in Fig. \ref{fig:flowchart}.

Previous studies integrating wind energy into ancillary markets have focused mainly on turbine-level control, which is commonly used for primary control. 
While substantial research has been conducted on wind energy participation in the FCR market (relevant to individual turbine performance, e.g. \cite{kayedpour_optimal_2023, hosseini_advanced_2020}), the secondary market, particularly with respect to turbine-to-turbine interactions, remains under-explored. 
Expanding on this body of work, the focus of this study shifts to farm-level control, where FRR is more relevant due to its timescale.
In this context, wake interactions play a critical role, as changes in turbine operation affect downwind turbines with a delay caused by wake interactions.
An FRR-providing WF aims to track a power reference signal generated by the TSO over a time interval of several minutes \cite{boersma_constrained_2019}.
This introduces the challenge of aerodynamic coupling between turbines within the farm, occurring at timescales comparable to the reference signal \cite{shapiro_modelbased_2017}.
For example, when the wind direction changes, it alters the effective layout of the WF and the interactions between turbines. 
Given typical turbine spacing (on the order of a kilometre) and wind speeds (around 10 \si[per-mode=symbol]{\meter\per\second}), it takes approximately two minutes (depending on conditions) for a downwind turbine to experience the wake effect of an upwind turbine. 
This underscores the importance of understanding wake interactions when participating in the secondary market, and control strategies that fail to account for wake advection time and wake interactions are likely to be inefficient or unsuccessful \cite{starke_yaw-augmented_2023}.

The proposed model in this paper is assessed through a case study of an actual offshore WF participating in the GB electricity markets, chosen for its high wind penetration.
GB has a significant share of wind power in its electric system, ranking second only to gas (approximately 20\% and 30\% respectively in 2018), but faces challenges with congestion and limited interconnections with neighbouring countries \cite{edmunds_participation_2019}.
Alongside Denmark and Ireland, the substantial levels of wind generation within these grids can serve as a benchmark for other power systems integrating larger shares of wind energy \cite{cole_critical_2023}.
The National Energy System Operator (NESO) aims for carbon-free operation of the GB transmission system by 2025 \cite{nedd_operating_2020}.
The impact of increasing penetration of wind generation on FCR requirements in GB has been analysed in \cite{vogler-finck_evolution_2015} and successful frequency response tests at the Burbo Bank offshore WF in Liverpool Bay have demonstrated the potential for wind power to support grid stability in GB \cite{noauthor_orsted_2016}.
In the GB power system, reserve services are procured by NESO through specific markets.
FCR is procured through Frequency Response, and depending on their size and location, power stations, including WFs (as large generators), may be required to provide this capability, known as Mandatory Frequency Response (MFR) \cite{noauthor_national_2024,edmunds_participation_2019}. 
The primary equivalent of FRR in GB is an upward-only service known as Fast Reserve (FR) \cite{edmunds_participation_2019}.
Throughout this paper, MFR and FR will be used as equivalents for these respective services\footnote{NESO is in the process of developing frequency response and reserve services: https://www.neso.energy/industry-information/balancing-services. This study focuses on established markets with available historical data.  }.
Although this study focuses on the GB reserve services markets, the presented methodology  can be applied to similar power systems and markets.
It is expected that the core findings of this study, which emphasise the importance of incorporating wake effects in WF participation in electricity markets, will remain consistent when applied to other power systems with different market structures.

\section{Scenario generation and reduction}
\label{scenarios}
Scenario generation for uncertain parameters is carried out using statistical models of FR demand, wind speed, and wind direction, these two latter enabling WF power calculation considering wake effects with the FLORIS simulation software.

\subsection{Scenario generation process}
\label{scenario generation}
Each scenario represents a specific hour in the day that corresponds to one outcome of the following uncertain parameters:
\begin{itemize}
    \item FR duration
    \item Wind speed
    \item Wind direction
\end{itemize}
The scenario generation process with respect to the above uncertain parameters is described in the following subsections.

\subsubsection{FR duration scenario generation}
\label{FR generation}
In order to generate random durations of FR activation, its statistical properties are analysed. 
The potential \% of instruction and number of instructions by FR duration from 1st April 2016 to 1st June 2021 is obtained from NESO \cite{noauthor_national_2024-2}.
The cumulative distribution function (CDF) is extracted from the data, to be compared with a random number from a uniform distribution.
When this number is greater than the CDF, it is rounded to the closest corresponding element, to model a random FR duration.

\subsubsection{Wind condition scenario generation}
\label{wind generation}
To generate wind condition scenarios, namely wind speed and wind direction, hourly average and its corresponding standard deviation has to be computed for both parameters. 
Once this is done, these can be used in probability distribution functions (PDF) to generate random values.
The PDF used to generate wind speed scenarios is a normal distribution, with $\mu$ being the forecasted average wind speed, and $\sigma$ its corresponding standard deviation.
The normal distribution is adopted in this study to represent the uncertainty associated with forecast errors in wind speed, rather than to describe the statistical distribution of wind speed itself.
Wind conditions for the following day are assumed to be forecasted in advance, with the corresponding uncertainty modelled as a normally distributed forecast error.
Since the formulation deals with short-term wind speed forecasts rather than long-term historical variability, the normal distribution is well suited for this framework, as forecast errors and short-term deviations are typically modelled as normally distributed processes.
While the normal distribution provides a common and practical approach for error estimation \cite{hodge_wind_2012} , particularly in wind power forecasting applications, more accurate or problem-specific error distributions have also been proposed in the literature \cite{hodge_wind_2012}.
However, at the scales considered in this study, the impact of employing a more sophisticated distribution is expected to be minor, and such refinements could be explored in future work. 
For wind direction, a normal distribution cannot be used, since the domain of the normal distribution could generate scenarios with wind directions greater than \ang{360}.
Instead, the circular analogue of the normal distribution, the von Mises distribution \cite{mardia_directional_2000}  is employed.
The circular average wind direction is denoted here using the parameter $\mu$, and the parameter $\kappa$ represents the concentration measure, so 1/$\kappa$ is analogous to $\sigma^2$, where $\sigma$ is the circular standard deviation.

\subsubsection{Number of scenarios}
\label{scenario numbers}
\begin{figure}[t] 
\centering
\includegraphics{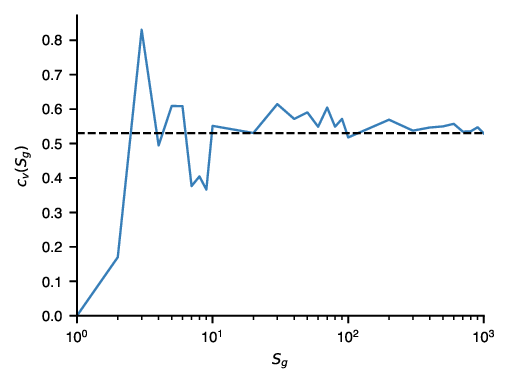}
\caption{The coefficient of variation of $E_{gen}$ as a function of the number of generated scenarios $S_g$}\label{fig:cv}
\end{figure}

In this study, we use the Energy Generated $E_{gen}$ as a mean estimator, to determine a sufficient value of the number of generated scenarios. 
For $S_g$ generated scenarios, $E_{gen}(S_g)$ is estimated as follows:
\begin{equation}
\label{Egen}
E_{gen}(S_g) = {\frac{1}{S_g}}\sum_{s=1}^{S_g}P_{gen}(w_s^s,w_d^s)\cdot \Delta t_{fr}^s
\end{equation}
where $P_{gen}$ is the generated WF power computed by FLORIS at wind speed $w_s^s$ and wind direction $w_d^s$, and $\Delta t_{fr}^s$ the FR duration, in the s-th scenario.
Therefore, $E_{gen}$ is only an index representing stochastic variables of FR duration and wind conditions.  
The coefficient of variation $c_v$ of $E_{gen}$ as a mean estimator is computed as follows:
\begin{equation}
\label{cv}
c_v(S_g) = \frac{\sigma_{E_{gen}}}{E_{gen}}
\end{equation}
In order to choose an accurate number of generated scenarios $S_g$, $c_v$ must be as stable as possible in order to ensure consistency. 
Fig. \ref{fig:cv} illustrates the evolution and convergence of $E_{gen}(s)$ as a function of the number of generated scenarios.
The figure exhibits significant variability for fewer than 10 scenarios, which is primarily due to the inherent randomness of the process.
The computation of $P_{gen}$  requires executing the FLORIS optimisation routine for each wind speed and direction scenario, with the computational time scaling proportionally with both the number of turbines in the WF and the total number of scenarios. 
For a large WF, such as the one considered in the case study presented in Section \ref{results and discussion}, running 1000 scenarios requires several hours. 
Given the exponential increase in wall-clock time, $c_v$ calculations beyond 1000 scenarios have not been performed. 

\subsection{Scenario reduction process}
\label{scenario reduction}
Although $c_v$ in Fig. \ref{fig:cv} stabilises much earlier than 1000 scenarios, this number is selected as it provides greater stability while maintaining a reasonable computational cost.
The aim of the scenario reduction process is to cluster the generated scenarios into $S$ representative scenarios.
Afterwards, the optimisation problem is solved considering the $S$ representative scenarios that compromises between computation time and result accuracy.
The K-medoids algorithm \cite{park_simple_2009} is used to cluster the scenarios.
To ensure that each uncertain variable has the same weight in scenario reduction, all of them are normalised, i.e. the input vectors are scaled individually according to their unit norm. 
Each cluster corresponds to a typical hour (representative scenario) that represents a set of similar hours.

\begin{figure}[t] 
\centering
\includegraphics{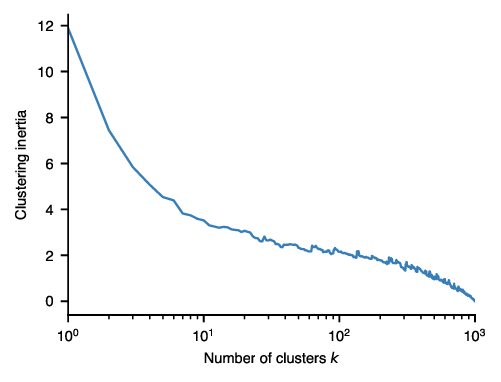}
\caption{Average within-cluster sums of point-to-medoid distances}\label{fig:inertia}
\end{figure}

The elbow method \cite{bholowalia_ebk-means_2014} is applied to find an appropriate value of representative scenarios $S$. 
It suggests selecting a number of clusters where the addition of another cluster results in minimal improvement of data modelling (scenarios).
In this respect, the K-medoids clustering algorithm is run on all $S_g$ generated scenarios for a range of values of $k\in\{1,...,S_g\}$, and for each value of $k$, the clustering inertia \cite{noauthor_scikit-learn-extra_2023} is computed.
Fig. \ref{fig:inertia} depicts the evolution of the clustering inertia as a function of $k$.
The knee point $k$ = 15 is chosen, where adding an extra cluster does not considerably reduce inertia.
Each cluster is represented by one scenario (medoid); therefore $S$ = 15 representative scenarios are considered in the optimisation problem formulated in Section \ref{optimisation}. 
For each cluster, the representative scenario is one of the original scenarios belonging to the cluster with minimum distance to the other scenarios of the cluster. 
The weight of each representative scenario in the optimisation problem (i.e. $\rho^s$), is equal to the number of scenarios belonging to the cluster over the total number of generated scenarios. 

\section{Optimisation process}
\label{optimisation}
The following problem proposes an optimal joint energy and reserve scheduling model for a WF to support decision-making for an optimal contribution to DA energy and reserve markets, applied to GB’s electric system.
In the proposed framework, the WPP is considered as a BRP (Balance Responsible Party), enabling its participation in the jointly cleared DA energy and reserve markets.
The WPP is also assumed to engage in two GB ancillary services, namely MFR and FR. 
Given that large WFs are required to provide MFR, the model incorporates contributions to upward MFR services, specifically low-frequency response categorised into primary and secondary response \cite{edmunds_participation_2019, noauthor_national_2024} , despite primary control not being the focal point of this study. 
Additionally, participation in FR is assumed, even though empirical feedback on WF engagement in this market remains unavailable at the time of study.
The optimisation is carried out for 24 hours, considering market parameters on an hourly basis.
This aligns with GB's wholesale electricity market, where DA hourly markets exhibit the highest trading volumes \cite{noauthor_epex_2024}.
This timescale is also relevant for MFR, as MFR payments follow a pay-as-bid structure, including hourly holding payments and response energy payments \cite{noauthor_national_2024}; however, the latter is not available for wind power \cite{nedd_operating_2020}.
While the settlement period for FR operates on a half-hourly basis, compensation is provided through hourly availability fees and utilisation payments \cite{noauthor_national_2024-2,nedd_operating_2020}.
To maintain consistent timescales, it is assumed that FR duration for one hour is the sum of the durations of the two half-hour periods.
This study focuses solely on the provision of upward reserve services, which present greater operational and economic challenges for wind power producers.
Unlike downward reserves, which can be provided at minimal cost by shutting down turbines, upward reserves require curtailing generation in advance, thereby sacrificing energy revenue to maintain headroom for under-frequency events \cite{edmunds_participation_2019}.
Similar modelling approaches that consider only upward reserve provision, due to the lack of fuel-saving benefits for wind power in downward regulation, have been adopted in prior studies \cite{soares_optimal_2016, liang_increased_2011}.

Under the proposed framework,  the WPP submits separate bids for energy and reserve markets for each time interval $t$ of the following day.
Scheduled bids are subject to real-time balancing, where deviations, caused by uncertainties in power generation, are financially settled through an imbalance pricing mechanism.
Both energy and reserve imbalances are settled in real time, where suppliers incur revenue reductions and penalties for deviations from their DA commitments. 
TSOs apply an imbalance pricing mechanism and require BRPs to compensate for positive and negative deviations at each settlement to ensure the real-time demand-supply balance at the system level. 
The specific imbalance settlement rules and penalties for deviations vary across markets \cite{bottieau_very-short-term_2020}. 
In GB, this process takes place within the Balancing Mechanism (BM) \cite{noauthor_national_2024-1}. 
For simplicity, hourly settlements are assumed.
The WPP is modelled as a price-taker \cite{soares_optimal_2016,liang_increased_2011,vilim_wind_2014,rahimiyan_evaluating_2011,soares_optimal_2017}, based on the assumption that the WF's generation capacity is relatively small compared to the total system generation, so that it does not affect market prices.

The WF's power output is determined using FLORIS, modelling wind speed and direction scenarios while applying wake steering to mitigate wake effects. 
The FLORIS optimisation routine identifies WT yaw angles that maximise power production based on wind speed, wind direction, and turbulence intensity.
The following problem has been formulated as a quadratic program\footnote{The developed code used in this publication is available in the public domain and can be accessed at the GitHub repository https://github.com/marinmf/optimal-wf-scheduling.} and solved with the IPOPT solver in the Pyomo Python package \cite{noauthor_ipopt_2020, noauthor_pyomo_2024,noauthor_pyomo_2024-1}.

\subsection{Objective function}
\label{objective}
The objective of this problem is to maximise the WPP's expected profit $\prod$ while minimising imbalance costs, taking into account the expected representative scenarios while satisfying their associated constraints.
The optimisation problem is defined for a dispatch time step of one hour, which will be solved for each 24 hours of the day. 
It is possible to solve each hour independently since there are no constraints between two time steps (i.e. there is no relation between the variables of each separate hour).
Therefore, for the set of scenarios (s $\in$ S), the optimisation problem for each time step $t$ is formulated as follows:
\begin{align}
\max_\phi \textstyle\prod = \ &P_e(t)\cdot\lambda_e(t) + P_{mfr}^\uparrow(t)\cdot\lambda_{mfr}^\uparrow(t) + P_{fr}^\uparrow(t)\cdot\lambda_{fr,a}^\uparrow(t) \nonumber\\
&+ \sum_{s=1}^{S}\rho^s\cdot[P_{fr}^\uparrow(t)\cdot\Delta t_{fr}^s(t)\cdot\lambda_{fr,u}^\uparrow \nonumber\\
&- ((P_{fr}^\uparrow(t) - \Delta P_{fr}^s(t))\cdot\Delta t_{fr}^s(t))^2\cdot\lambda_{b,fr}(t)^2 \nonumber\\
&- (P_e(t) - \Delta P_e^s(t))^2\cdot\lambda_{b,e}(t)^2]
\label{objective function}
\end{align}
where $\phi$ is the set of the decision variables of the optimisation problem as defined in the Nomenclature.
The first line represents the first stage of the stochastic programming formulation.
First stage optimisation variables $P_e$, $P_{mfr}^\uparrow$, and $P_{fr}^\uparrow$, are bidding decision variables of energy production and upward reserves.
This stage includes revenue of selling energy in the wholesale market (first term) and revenue from selling upward reserves (second and third terms).
The following lines represent the second stage of the stochastic programming formulation.
Second stage optimisation variables $\Delta P_{fr}^s$ and $\Delta P_e^s$, are real-time upward reserve and energy re-dispatches.
This stage includes revenue of FR energy deployment (second line), cost of FR energy imbalances (third line), and cost of energy imbalances (fourth line). 
Imbalance and penalty terms are squared to ensure positivity and adopt a conservative approach, as this formulation compels the identification of the closest solution.

\subsubsection{Constraints}
\label{constraints}
The objective function (\ref{objective function}) is subject to the following constraints:
\begin{equation}
\label{c1}
P_e(t) + P_{mfr}^\uparrow(t) + P_{fr}^\uparrow(t) \leq P_{max}(t), \forall t
\end{equation}
\begin{equation}
\label{c2}
P_e(t) \geq P_{min}, \forall t
\end{equation}
\begin{equation}
\label{c3}
P_{mfr}^\uparrow(t) \leq 0.1\cdot P_e(t), \forall t
\end{equation}
\begin{equation}
\label{c4}
P_{fr}^{\uparrow}(t) \geq P_{fr}^{min\uparrow}, \forall t
\end{equation}
\begin{equation}
\label{c5}
P_e(t), P_{mfr}^\uparrow(t), P_{fr}^\uparrow(t) \geq 0, \forall t
\end{equation}
\begin{equation}
\label{c6}
\Delta P_e^s(t) + P_{mfr}^\uparrow(t) + \Delta P_{fr}^s(t) \leq P_{max}^s(t), \forall t, \forall s
\end{equation}
\begin{equation}
\label{c7}
\Delta P_e^s(t), \Delta P_{fr}^s(t)  \geq 0, \forall t, \forall s
\end{equation}
Constraint (\ref{c1}) limits the scheduled energy and upward reserve contributions to the WF’s maximum forecasted capacity, defined by wind speed and wind direction averages of the hour.
Constraint (\ref{c2}) limits energy contribution to WF minimum capacity, considered at 0 MW due to the absence of mechanical constraints that prevent its complete shutdown.
Constraint (\ref{c3}) limits upward MFR contribution to 10\% of the energy contribution, which represents a 10\% deloading strategy. 
This strategy is applied in NESO’s MFR service, where WFs are required to hold a 10\% reserve of power output while in Frequency Sensitive Mode \cite{cole_critical_2023}. 
Furthermore, \cite{ahmadyar_coordinated_2017} found that the deloading margin for an individual WT should not exceed 10\%.
Constraint (\ref{c4}) ensures that FR delivery is larger than the minimum requirement for this service. 
In the GB market, the minimum requirement $P_{fr}^{min\uparrow}$ is 25 MW \cite{noauthor_national_2024-2}.
Constraint (\ref{c5}) ensures positivity of first stage decision variables. 
Constraint (\ref{c6}) restricts upward reserve and real-time energy redispatches to WF maximum available output power for each scenario, using representative scenarios of wind speed and wind direction.
Constraint (\ref{c7}) ensures positivity of second stage decision variables.

The objective function in (\ref{objective function}), along with constraints (\ref{c1})-(\ref{c7}), formulates the two-stage stochastic optimisation problem designed to maximise the WPP's profit in the energy and reserve markets.
A distinguishing feature of the proposed methodology is the explicit consideration of wake effects, along with the implementation of a mitigation strategy aimed at improving WF performance and market participation in short-term electricity markets. 
To highlight the improvements achieved, the proposed model is compared with the previously introduced "Power Curve" and "Baseline" approaches. 
The methodology employs wake steering as a mitigation technique, referred to as "WRC", which is a short form of Wake Redirection Control.
All three approaches are applied to the case study in Section \ref{results and discussion}, where computation of maximum output powers $P_{max}$ and $P_{max}^s$ in FLORIS is performed differently for each approach. 
In the proposed framework, wake effects are incorporated through the computation of the velocity field and wake interactions in FLORIS.
In the Power Curve approach, the WF is modelled in FLORIS without applying a wake model, assuming uniform inflow conditions (i.e. no wake in the flow).
In contrast, the Baseline approach includes wake modelling within FLORIS, while the WRC approach further optimises turbine yaw angles to mitigate wake losses and enhance overall WF performance. 
The additional power generated through wake steering can be allocated using two distinct strategies.
The first strategy optimally distributes the power increase between energy and reserve contributions to maximise economic efficiency. 
This strategy, denoted by "WRC1", maintains the same constraints as the "Power Curve" and "Baseline" approaches. 
However, as wake steering is still a relatively new control strategy, its continuous application may not always be optimal due to concerns from WF operators about potential increases in turbine loading resulting from intentional yaw misalignment \cite{houck_review_2022}.
Therefore, the second strategy, denoted "WRC2", allocates the additional power gained by wake steering exclusively to upward reserves, rather than the energy market, when it exceeds the maximum power achievable with non-yawed turbines, as presented in \cite{boersma_constrained_2019}.
To enforce this condition, an additional constraint is introduced specifically for the WRC2 strategy, formulated as follows: 
\begin{equation}
\label{c8}
P_e(t) \leq P_{max}^{Baseline}(t), \forall t
\end{equation}
and the power limit in (\ref{c1}) is $P_{max}^{WRC}$.

\subsubsection{Income}
\label{income}
Based on the hourly bids submitted to the DA energy and reserve markets, the expected hourly income $\prod_h$ is determined using the following equation: 
\begin{align}
\textstyle\prod_h = \space & P_e(t)\cdot\lambda_e(t) + P_{mfr}^\uparrow(t)\cdot\lambda_{mfr}^\uparrow(t) + P_{fr}^\uparrow(t)\cdot\lambda_{fr,a}^\uparrow(t) \nonumber\\
&+ \sum_{s=1}^{S}\rho^s\cdot[P_{fr}^\uparrow(t)\cdot\Delta t_{fr}^s(t)\cdot\lambda_{fr,u}^\uparrow \nonumber\\
&- |(P_{fr}^\uparrow(t) - \Delta P_{fr}^s(t))\cdot\Delta t_{fr}^s(t)|\cdot\lambda_{b,fr} \nonumber\\
&- |P_e(t) - \Delta P_{e}^s(t)|\cdot\lambda_{b,e}]
\label{hourly income}
\end{align}
The first line represents the revenue from selling energy in the wholesale market, the MFR holding payment, and the FR availability payment.
The second line incorporates revenue from the FR utilisation payment, while the third and fourth lines account for the cost of FR and energy imbalances, respectively.
The total expected daily income $\prod_d$ is then computed by summing the hourly incomes:
\begin{equation}
\textstyle\prod_d = \displaystyle\sum_{h=0}^{23}\textstyle\prod_h
\label{daily income}
\end{equation}
These calculations are performed using the specific variables generated by the optimisation algorithm for each approach.

\section{Case study and simulation results}
\label{results and discussion}
This section presents the results of a case study focusing on the participation of the London Array WF.
This WF was selected as a representative case study. 
Although similar overall trends are expected for other offshore WFs operating under comparable conditions, we do not assert that the quantitative results are directly generalisable. 
The outcomes are influenced by various factors, including site-specific WF characteristics, atmospheric conditions, and electricity market pricing. 
The primary objective of this study is to demonstrate the feasibility of integrating wake modelling into market participation optimisation frameworks. 
The London Array, with a capacity of 630 MW, is situated 20 km off the Kent coast in the UK, as detailed in \cite{lanzilao_new_2022}.
This analysis evaluates its participation in various GB markets, including the DA energy market, upward MFR (Primary and Secondary responses), and FR markets.
The proposed operational strategy is assessed through a structured evaluation process.
First, the optimisation problem described in Section \ref{optimisation} is solved for two specific days, as outlined below.
Subsequently, the results of each approach presented in Section \ref{research gap} are analysed to assess their respective effectiveness.
Finally, the implications of the findings and key takeaways are discussed.

\subsection{Case description} 
\label{case study}
We do not have access to the exact wind conditions on the London Array WF site.
Instead, we use the wind dataset obtained from an offshore meteorological mast near Ijmuiden, Netherlands \cite{noauthor_tno_2016}.
It comprises wind speed and direction measurements recorded at a frequency of 4 Hz at a height of 90 m, corresponding to the hub height of the turbines in the London Array WF, and spans from January 1st to December 31, 2015.
It should be noted that using measurements from a different site introduces a degree of uncertainty due to potential differences in local wind conditions. 
In practice, such data must be derived from wind condition forecasts, which can be generated using methods such as ARIMA or various machine learning techniques, among others \cite{liu_short-term_2021}.
For each month, the average sensor readings at each interval were computed and subsequently resampled to an hourly time interval, computing the hourly means and standard deviations of wind speed and direction.
It is important to note that for wind direction, usual arithmetic mean and standard deviation are unsuitable for angular data.
This limitation arises because \ang{0} and \ang{360} represent the same angle (i.e. equivalent modulo a full cycle), making \ang{180} an invalid mean for angular values such as \ang{1} and \ang{359}.
To address this, circular statistics were employed to compute the circular mean and circular standard deviation \cite{mardia_directional_2000}.
Similar resampling for hourly means has previously been  applied for wind speed and wind direction \cite{liu_short-term_2021,tyass_wind_2022,yatiyana_wind_2017}.

\begin{figure}[t] 
\centering
\includegraphics{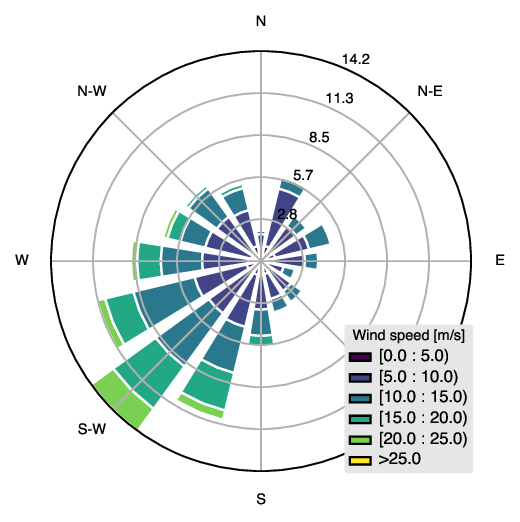}
\caption{Wind rose for the dataset with normed (displayed in percent) results}\label{fig:windrose}
\end{figure}

\begin{figure}[t] 
\centering
\includegraphics{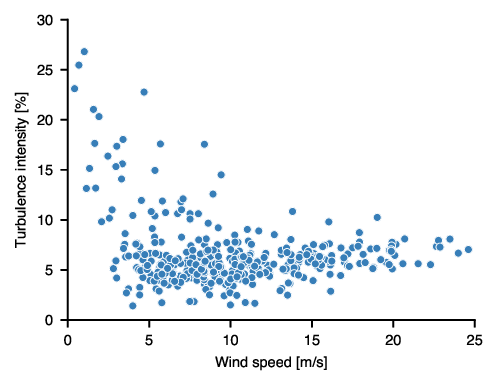}
\caption{Turbulence intensity as a function of wind speed for 10-minute intervals of all days in 2015}\label{fig:ti}
\end{figure}

\begin{figure}[t] 
\centering
\includegraphics{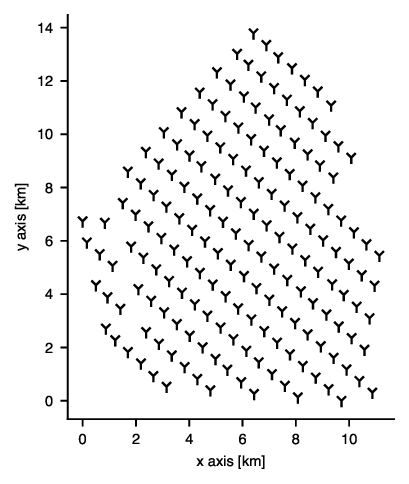}
\caption{Layout of the London Array farm}\label{fig:layout}
\end{figure}

The London Array WF was selected primarily because it is a real WF located in GB and because the wind dataset used in this study aligns with its wind conditions.
Fig. \ref{fig:windrose} depicts the wind rose for the dataset, showing the distribution of wind speed and direction.
This distribution closely resembles the mast data from the London Array \cite{watson_report_2017}, with a predominant south-Westerly wind direction.
Fig. \ref{fig:ti} illustrates the turbulence intensity (TI) of the dataset as a function of wind speed, where TI values were computed over 10-minute intervals from 12:00 to 12:10 for all days in 2015.
The exponential decay aligns with findings in the literature \cite{nygaard_wakes_2014}, which also included the London Array.
For the WF representation in FLORIS, turbine locations were extracted from the Operations Brochure \cite{london_array_limited_london_2020}.
Since turbine coordinates were provided in the WGS 84 geodetic system, they were transformed into the UTM coordinate system to enable representation on a Cartesian plane.
The easting coordinate was mapped to the $x$-axis, and the northing coordinate to the $y$-axis.
The resulting WF layout is shown in Fig. \ref{fig:layout}.
For the remainder of the study, the westerly wind direction (\ang{270}) is aligned with the $x$-axis, while the $y$-axis corresponds to the southerly wind direction (\ang{180}), consistent with the approach in \cite{lanzilao_new_2022}.

\begin{figure}[t] 
\centering
\includegraphics[width=\textwidth,keepaspectratio]{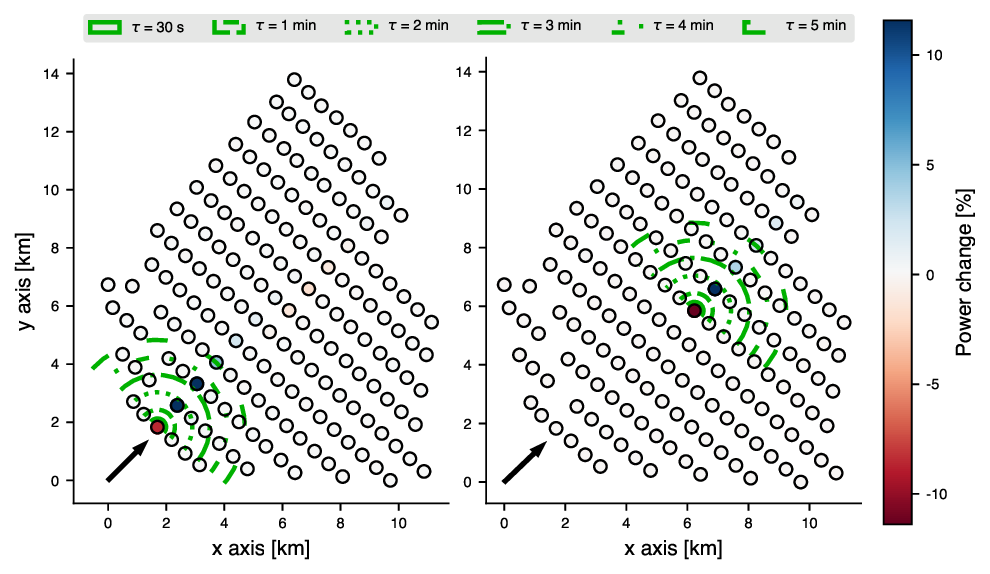}
\caption{Power variation of WTs in the WF after yawing a turbine. Circles represent the distance travelled by the wind after the corresponding time intervals, and the arrow indicates the incoming wind direction. The parameter $\tau$ denotes the time required for the wake of the yawed turbine (in red) to propagate downstream.}\label{fig:power_variation}
\end{figure}

As discussed in Section \ref{methods},  the response of downwind turbines to operational changes in the incoming wind, caused by wind direction changes or the implementation of wake steering, exhibits a natural latency due to the time required for upwind turbine wakes to be advected downstream.
It is important to ensure that the wake propagation and this latency remains sufficiently short to ensure effective participation in secondary control.
Simulations were performed using FLORIS to assess the compatibility of the timescale of wake propagation with FR timescales.
The power output of each turbine in the WF was computed under a wind speed of 10 \si[per-mode=symbol]{\meter\per\second} and a wind direction of \ang{225}.
Subsequently, a single turbine in the farm was yawed by \ang{20} under the same wind conditions, and the power outputs of all turbines were recomputed.
The difference in power output between the two cases was determined for each turbine and shown in Fig. \ref{fig:power_variation}.
Results are shown in Fig. \ref{fig:power_variation} for two scenarios : (i) yawing the turbine in the front row, and (ii) yawing the turbine in the middle of the farm.
Green circles in the figure indicate the estimated time required for each downwind location to detect the influence of a yaw angle change in the upwind turbine, while the turbine located at the centre of the circle is shown in red, as it experiences reduced power production when yaw misalignment (wake steering) is applied.
This time estimate is based on a simplified assumption that the wake propagates downstream at the incoming wind speed, neglecting spatial velocity variations within the WF.
According to \cite{noauthor_national_2024-2}, FR active power delivery must begin within two minutes of the dispatch instruction and must be sustained for a minimum of 15 minutes.
Fig. \ref{fig:power_variation}  illustrates that the increase in power output of the first downwind row (caused by intentionally yawing the upwind turbine) occurs within approximately two minutes, with the furthest observable effects propagating 2-3 rows downstream in under five minutes, which is within timescales relevant to the secondary market.
Therefore, by the time FR active power delivery is initiated, the WF will already have achieved a noticeable portion of the power output enhancement realised by wake steering. 
Fig. \ref{fig:power_variation} demonstrates that wake steering provides a timely response suitable for secondary market participation.
Two important observations are highlighted: 
\begin{itemize}
    \item When a turbine is deliberately yawed, there is a delay before downwind turbines experience the resulting wake deflection, primarily determined by turbine spacing and wind velocity.
    \item Yawing a turbine predominantly alters the power output of local neighbouring turbines (mainly the first two downwind rows), enabling relatively fast redistribution of power within the WF. 
\end{itemize}
It is important to note that we employ a steady-state engineering wake model to represent wake effects which are inherently unsteady.
Moreover, our wake propagation timescales shown in Fig. \ref{fig:power_variation} are derived from a highly simplified assumption, namely that the wake travels downstream at the incoming wind speed.
While this approach offers computational efficiency, it does not capture the unsteady dynamics introduced by changes in atmospheric conditions (e.g. wind direction) or turbine yaw angles.
More accurate representations could be achieved using dynamic wake models such as those proposed in \cite{becker_ensemble-based_2022,becker_dynamic_2025,starke_dynamic_2024}, which account for these transient effects.
However, the added complexity and computational demands of such models can make them impractical for optimisation studies focused on market participation.
Despite the simplicity of our steady-state approach, it provides valuable insight and confirms that the benefits of wake steering can be seen within the required time-frame for secondary market participation.

\subsection{Results and Discussion}
\label{results}
Simulations were conducted on 11 and 12 April 2015, selected for the following reasons:
\begin{itemize}
    \item Wind speeds on these days fall within Regions 2 and 3 of the WTs (rated wind speed of 14 \si[per-mode=symbol]{\meter\per\second}) at London Array.
    \item The standard deviation of wind direction is moderate, representing days with notable, yet common, wind variations.
    \item The direction of the wind ranges between \ang{-50} and \ang{-150}, which is typical for the site (refer to Fig. \ref{fig:windrose}). 
However, this range has the potential to produce strong wake effects (particularly near \ang{-90}), according to Fig. \ref{fig:layout}.
\end{itemize}
WTs in the farm are modelled in FLORIS using a Turbine object configured with the SWT-3.6-120 characteristics from \cite{bauer_siemens_2017}.
Parametric inflow conditions are an air density of 1.225 \si[per-mode=symbol]{\kilo\gram\per\cubic\meter} and a wind shear exponent of 0.12 (FLORIS input file default values), and WT yaw angles are constrained between \ang{+-25} to prevent excessive structural loading on turbines under significant yaw misalignment \cite{boersma_constrained_2019}.
To optimise yaw angles for maximising WF power production through wake steering, the Geometric Yaw optimiser presented in \cite{stanley_enabling_2023} is used. 
This method offers a balance between computational efficiency and result accuracy, leveraging WF geometry to quickly determine approximately optimal yaw angles.

The input data for the optimisation problem includes hourly datasets for wind speed and direction (as previously described), TI, and market prices.
Hourly TI was computed by averaging 10-minute TI values over each hour. 
Given that all market prices appear linearly in (\ref{objective function}) and that the WPP is assumed to be a price taker, market price uncertainties are replaced with their expected values \cite{soares_optimal_2016}.
Energy market price data corresponds to DA spot prices for GB on April 11 and 12, 2015, sourced from \cite{noauthor_open_2017}.
The MFR holding payment is calculated as the weighted average of each BM Unit's Primary and Secondary response holding bids, weighted by their respective holding volumes for April 2015 \cite{noauthor_national_2024}.
Market information for FR (availability and utilisation payments) was obtained for April 2019 due to the unavailability of prior data for that month \cite{noauthor_national_2024-2}.
At the time, FR was procured on a monthly basis \cite{edmunds_participation_2019}, making relative payments constant parameters in the optimisation problem.
This is no longer the case, as FR is now procured within day through the Optional FR service \cite{noauthor_national_2024-2}; however, no information regarding this new market is currently available.
The utilisation fee used was the monthly average, while the availability fee was computed as the average of all fees divided by their contracted volumes.
Penalties for energy and FR imbalances were set at 20\% of their spot and utilisation prices, respectively.
The proposed optimal joint energy and reserve scheduling model is executed on a personal computer with Intel Core i7 CPU 2.3 GHz processor.
The solution for the optimal schedule took 30 min for each day of the case study.

\begin{figure}[t] 
\centering
\includegraphics[width=\textwidth,keepaspectratio]{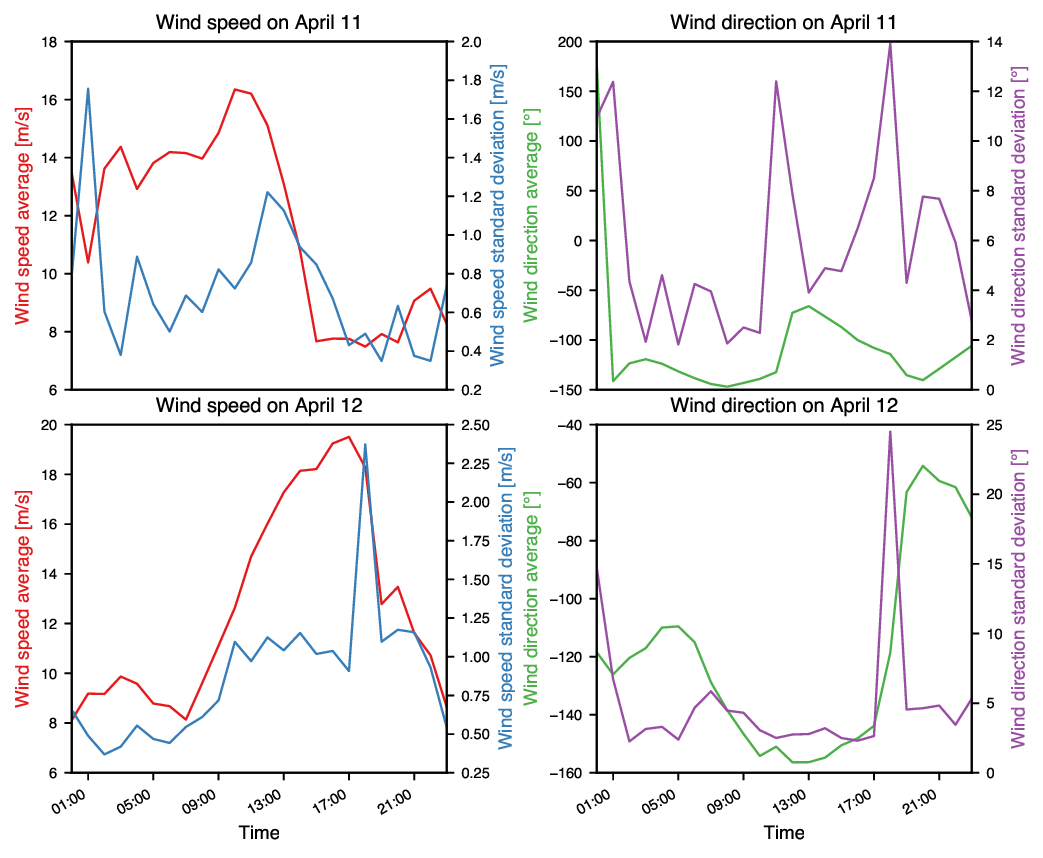}
\caption{Averages and standard deviations of wind speed and direction on April 11 and 12, 2015}\label{fig:avg_std}
\end{figure}

Fig. \ref{fig:avg_std} illustrates averages and standard deviations of wind speed and direction on April 11 and on April 12.
The wind speed on the 11th exhibits an increasing trend in the early hours, peaking near midday, followed by a decline to relatively low levels later in the day.
The wind direction on that day undergoes a complete shift within the first two hours, stabilising at approximately \ang{-100} for the remainder of the day.
On the 12th, wind speed is initially low during the early hours, gradually increasing to reach its peak at 17 h, before declining again later in the day.
The direction of the wind on that day initially fluctuates around \ang{-120}, trends toward \ang{-160} by midday, and then undergoes a sharp shift to approximately \ang{-60} by the end of the day.

\begin{figure}[t] 
\centering
\includegraphics[width=\textwidth,keepaspectratio]{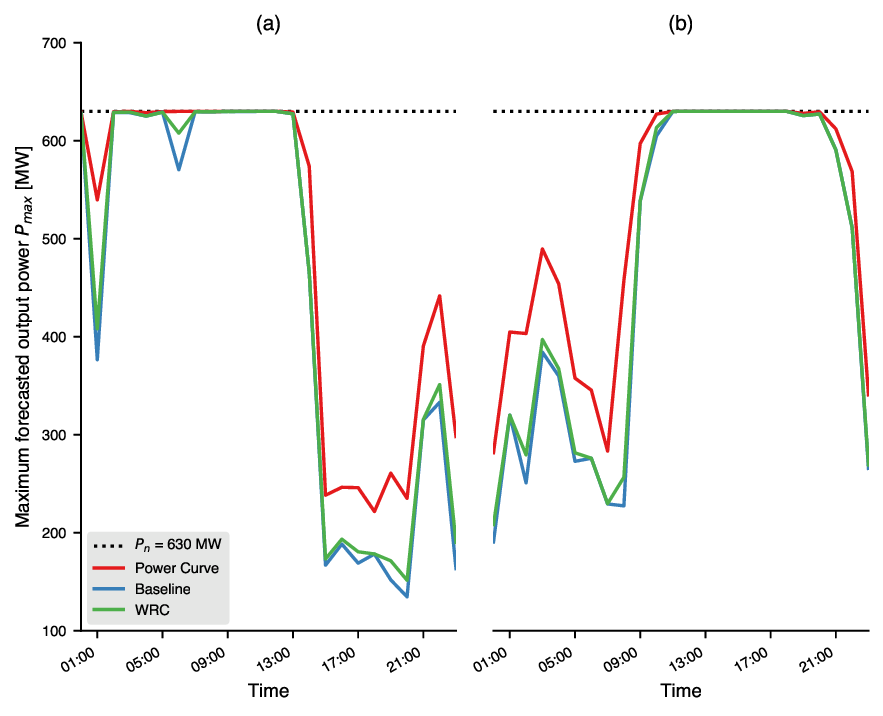}
\caption{Maximum forecasted output power $P_{max}$ for April 11 (a) and April 12 (b), 2015}\label{fig:pmax}
\end{figure}

Fig. \ref{fig:pmax} presents the WF's maximum forecasted output power $P_{max}$ for both days as determined by each approach discussed in Section \ref{constraints}.
Fig. \ref{fig:pmax} shows that output power is at the nameplate (i.e. rated) capacity of the WF (630 MW) from 02:00 to 13:00 on the 11th and from 11:00 to 20:00 on the 12th, and there is minimal difference between the approaches. 
This is because, as shown in Fig. \ref{fig:avg_std}, the wind speed is sufficiently high during this period that the turbines operate in Region III, even when wake effects are considered, so regardless of the modelling approach used, the turbines are predicted to produce their rated power. 
However, when wind speeds are lower and turbines more frequently operate in Region II (e.g. after 13:00 on the 11th), the Power Curve approach consistently yields the highest values, as it neglects wake effects that lead to power losses.
These values can exceed those of the Baseline approach by up to 82\% on the 11th and 101\% on the 12th, underscoring the critical importance of accounting for wake effects.
It is also noteworthy how significantly wind direction influences power production. 
On April 11, the Baseline approach predicts lower power output at 06:00 compared to 05:00, despite higher wind speeds, due to a less favourable wind direction.
This once again highlights the sensitivity of WF power production to wind direction due to wake effects and reinforces the need for incorporating wake effects in our modelling.
Reference \cite{porte-agel_numerical_2013} showed that even small changes in wind direction can substantially affect WF power output, where a \ang{10} change in wind direction can increase total power output by as much as 43\%.
The use of wake steering generally improves WCR values (i.e. both WRC1 and WRC2 approaches, as they share the same $P_{max}$ ) over Baseline ones, with improvements of up to 16\% on the 11th and 13\% on the 12th, highlighting its usefulness.

\begin{figure}[t] 
\centering
\includegraphics[width=0.9\textwidth,keepaspectratio]{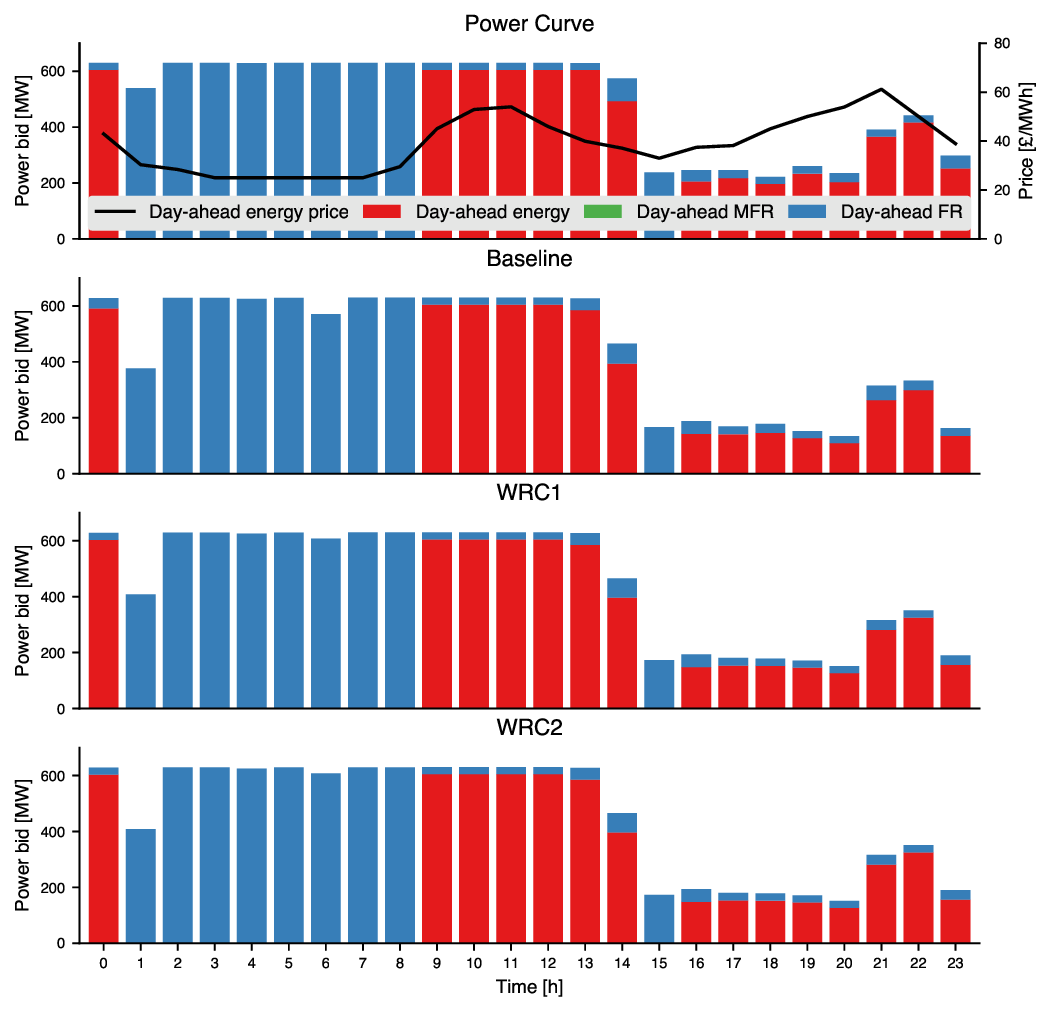}
\caption{Estimated scheduled energy and reserve contributions on April 11, 2015, for Power Curve, Baseline, WRC1, and WRC2 approaches}\label{fig:april_11}
\end{figure}

\begin{figure}[t] 
\centering
\includegraphics[width=0.9\textwidth,keepaspectratio]{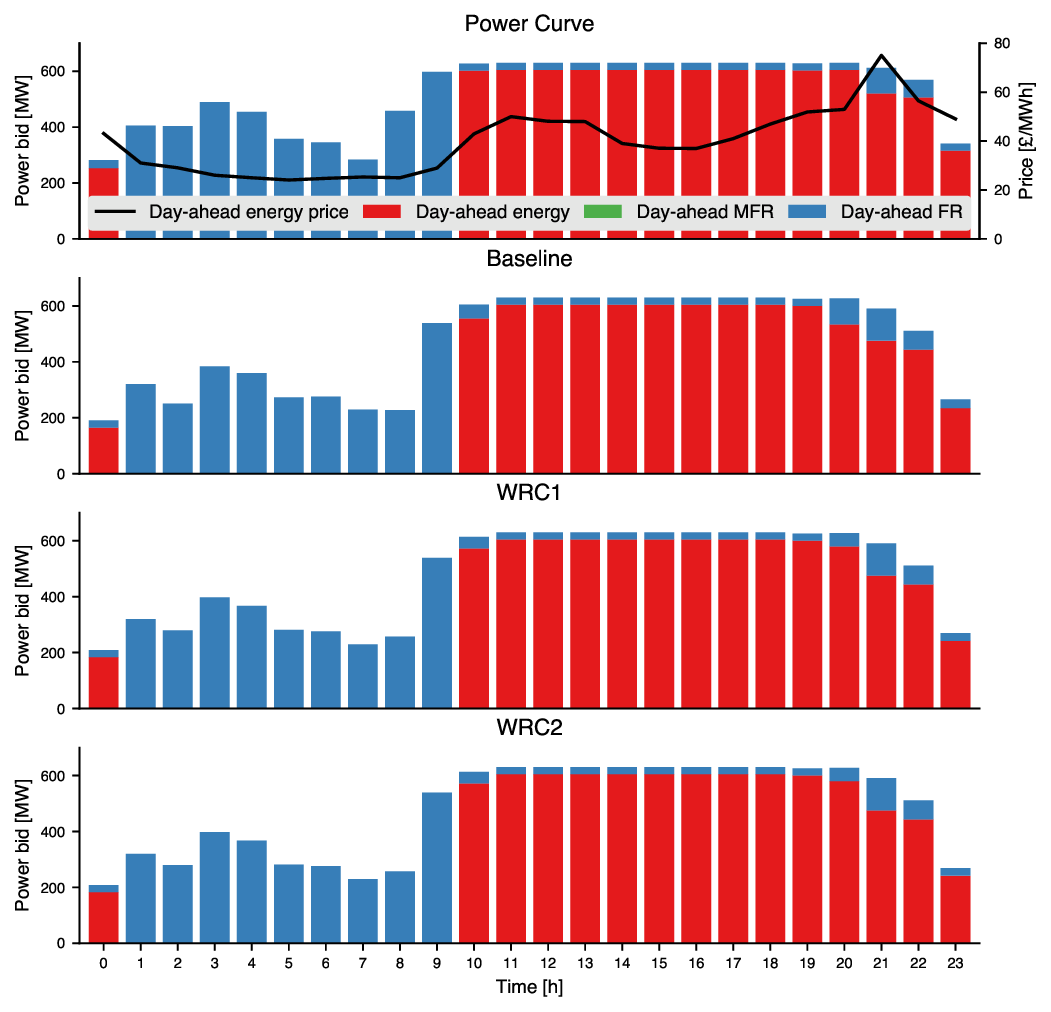}
\caption{Estimated scheduled energy and reserve contributions on April 12, 2015, for Power Curve, Baseline, WRC1, and WRC2 approaches}\label{fig:april_12}
\end{figure}

The hourly bids submitted by the WPP in the DA energy and reserve markets for each approach on both days are shown in Fig. \ref{fig:april_11} and Fig. \ref{fig:april_12}, respectively.
The Power Curve approach exhibits the highest bids in the energy and FR markets in Fig. \ref{fig:april_11} and Fig. \ref{fig:april_12}, since its maximum forecasted capacity in constraint (\ref{c1}) is greater.
Bids in the MFR market are consistently 0 MW for all approaches, suggesting that participation in this market is not economically attractive.
This observation aligns with the current state of the MFR market, where most WFs are still unwilling to participate and price themselves out of the market \cite{edmunds_participation_2019}.
On both days, the energy price decreases during the early hours, reaching its minimum in the morning, followed by a sharp increase towards noon. 
In the afternoon, the price declines once more before rising again in the evening, peaking at 21:00.
To evaluate the impact of this variation on the optimal solution,  a sensitivity analysis was conducted by varying energy prices, FR utilisation payments, and imbalance penalties under different cases of wind conditions.
The first case was a real hour (09:00 on 16/01/2015) with stable wind conditions, characterised by an average wind speed of 14.3 \si[per-mode=symbol]{\meter\per\second} ($\sigma$ = 0.8 \si[per-mode=symbol]{\meter\per\second}) and wind direction of \ang{-113.8} ($\sigma$ = \ang{3.3}).
The second case was a fictional hour characterised by high wind variability, featuring an average wind speed of 12.4 \si[per-mode=symbol]{\meter\per\second} ($\sigma$ = 3.5 \si[per-mode=symbol]{\meter\per\second}) and wind direction of \ang{62.5} ($\sigma$ = \ang{80.5}).
Under stable conditions (first case), energy prices above approximately \pounds35/MWh consistently led the model to favour maximum energy market bids and minimum FR bids (25 MW, as constrained in (\ref{c4})), while lower prices shifted all bidding activity to the FR market. 
Variations in imbalance penalties related to energy and FR imbalances had no observable effect on bidding behaviour, while FR utilisation payments only influenced bidding once they exceeded a threshold of approximately \pounds95/MWh, below which the model tended to bid minimally in the FR market. 
In contrast, under variable wind conditions (second case), active FR market participation was observed even at lower utilisation payments ($\sim$ \pounds10/MWh), and imbalance penalties as low as 10\% of market prices triggered a shift toward reserve market bidding.
These results indicate that the optimisation algorithm is sensitive to wind variability and uncertainty, which can substantially influence bidding behaviour by increasing the attractiveness of the FR market, even during periods of high energy prices, particularly when imbalance penalties are considered in anticipation of potential deviations.
A corresponding figure illustrating these results is provided in \ref{app1}.
The monthly analysis of wind speed and direction in the dataset reveals that large deviations in wind direction primarily occur at lower wind speeds and are relatively rare.
Although the underlying meteorological causes contributing to these deviations remain unidentified, their impact on bidding behaviour is acknowledged, with the assumption that wind direction remains stable in the majority of cases. 

Fig. \ref{fig:april_11} demonstrates that between 01:00 and 08:00, as well as at 15:00, all bids (irrespective of the approach) are directed to the FR market, driven by an energy price below \pounds35/MWh.
Similarly, Fig. \ref{fig:april_12} shows this trend from 01:00 to 09:00.
Conversely, bids in the FR market are minimal, with maximum bids directed to the energy market from 09:00 to 13:00 in Fig. \ref{fig:april_11} and from 11:00 to 19:00 in Fig. \ref{fig:april_12}.
This behaviour is attributed to favourable energy prices, high wind speeds, and moderate wind direction standard deviations during these hours.
However, from 16:00 to 23:00 in Fig. \ref{fig:april_11} and from 10:00 to 11:00 and 20:00 to 23:00 in Fig. \ref{fig:april_12}, bids in the FR market are no longer minimal despite high energy prices.
This shift is explained by low wind speeds and relatively high wind direction standard deviations, indicating variable wind conditions during these periods.
The observed allocation of all available power to the FR reserve market when the DA energy price is low arises since the model assumes that the WF acts as a price taker.
This implies that individual imbalances, including large curtailments, do not influence the overall imbalance price, a reasonable assumption when the participant’s capacity is small relative to the total market size.
Under these conditions, allocating power to the reserve market maximises expected profits, and the model behaves rationally within its assumptions.
In real systems, however, large-scale participation of WFs in the FR market or correlated forecast errors across multiple sites could impact imbalance prices and increase curtailment risk.
Although imbalance penalties are included, the model’s simplified treatment may not fully capture the operational consequences of widespread reserve market participation.
Consequently, while the modelled strategy appears economically optimal for individual WFs, it may not align with system-level reliability requirements or long-term renewable integration goals. 
For larger market participants (price makers) or when forecast errors are correlated across multiple WFs , imbalance prices are likely to increase with the size of imbalance.
Incorporating a variable imbalance price model would therefore provide a more realistic representation of market feedback mechanisms and enable a more comprehensive assessment of curtailment risks and grid stability impacts. 

Figures \ref{fig:april_11} and \ref{fig:april_12} reveal that both WRC strategies yield identical results.
This occurs because the distinguishing constraint between them (see Equation (\ref{c8})) is never activated.
For this constraint to become active, the following conditions must be satisfied:
\begin{itemize}
    \item The power produced in the WRC approach must exceed that of the Baseline approach, i.e. wake steering must enable surplus production.
This depends on wind conditions, particularly the presence of strong wake effects.
    \item The energy price must exceed \pounds35/MWh; otherwise, all bids are directed to the FR market.
This threshold is specific to this case study, as the FR utilisation payment is treated as a constant parameter, while energy prices are variable.
It is based on the available data, and in a real context where FR payments are variable, this condition would not apply.
    \item The optimal energy bid must exceed the maximum output power achievable in the Baseline approach $P_{max}^{Baseline}$.
\end{itemize}
In the cases studied, there was no time at which all the above conditions were simultaneously satisfied, which explains why both WRC strategies yielded identical results.

Based on the hourly bids submitted to the DA energy and reserve markets illustrated in Fig. \ref{fig:april_11} and Fig. \ref{fig:april_12}, the hourly income for each approach is calculated using equation (\ref{hourly income}), and the total expected daily income for each approach using equation (\ref{daily income}).
However, since the Power Curve approach represents an upper limit of achievable power, the resulting expected income can be unrealistic. 
To estimate the \textit{real} income for the Power Curve approach, imbalances are calculated based on the difference between its bidding decision variables and the upward reserve and real-time energy re-dispatches of the Baseline approach, as these reflect the actual power delivered.

Table \ref{tab:income} summarises the expected incomes on each day of the case study for each approach, including the \textit{real} income for the Power Curve approach, referred here as "Real PC". 
\begin{table}[t]
\centering
\begin{tabular}{cccccc}
\toprule
\multirow{2}{*}{\centering Day} &
\multicolumn{5}{c}{Expected daily income $\prod_d$ [\pounds]} \\ 
\cline{2-6} & Power Curve & Baseline & WRC1 & WRC2 & Real PC \\
\midrule
11.04.15 & 487'046 & 438'248 & 446'130 & 446'130 & 423'742 \\ 
12.04.15 & 543'555 & 494'473 & 499'814 & 499'814 & 478'425 \\
\bottomrule
\end{tabular}
\caption{Expected daily incomes for all approaches}\label{tab:income}
\end{table}
Due to high power estimations, the Power Curve approach predicts the highest income, estimating earnings 11\% higher on the 11th and 10\% higher on the 12th compared to the Baseline approach.
However, in reality, the Power Curve approach (Real PC) earns 13\% less than its estimates on the 11th and 12\% less on the 12th, resulting in an actual income that is 3\% lower than the Baseline approach on both days.
The Baseline approach serves as the most appropriate reference for evaluating the Power Curve method, as it explicitly accounts for wake effects through physics-based modelling.
 Nevertheless, it does not represent an absolute ground truth, since it relies on predicted inflow characteristics, derived from statistical models of wind speed and direction, and on wake predictions obtained from engineering models. 
Any inaccuracies in these inputs inevitably limit its accuracy. 
Thus, the Baseline approach should be regarded as a physically informed benchmark that highlights the types of errors likely to occur when power generation is estimated solely from power curves. 
The actual deviation between modelled and real performance ultimately depends on the accuracy of both inflow and wake predictions. 
This outcome highlights that neglecting wake effects can lead to high estimations of potential gains, with additional imbalance costs incurred due to the Power Curve approach's inability to meet its power commitments.
These imbalance penalties, set at 20\% of market prices in this study, significantly reduce real income and would have an even greater impact if they were higher.
In the EU, balancing costs have increased due to the variability and uncertainty of wind energy, particularly in regions where wind penetration reaches up to 20\% of total energy demand.
In most Member States with wind power contributing more than 2\% to annual generation, wind power producers are balancing responsible in financial or legal terms \cite{ewea_balancing_2015}. 
Often, these producers already bear the additional balancing costs they incur \cite{ewea_balancing_2015}.
In certain instances, balancing costs may surpass the actual generation costs, posing significant obstacles to the deployment of new wind energy installations \cite{ewea_balancing_2015}.
This highlights the substantial impact of imbalance costs, emphasising the importance of obtaining the most accurate estimates of wind power production (i.e. considering wake effects) for its participation in short-term electricity markets.
Conversely, the use of wake steering in the WRC approach (both strategies) demonstrates a 2\% increase in income on the 11th and a 1\% increase on the 12th compared to the Baseline approach, underscoring its effectiveness as a wake management technique.

\subsection{Assumptions and Limitations}
\label{limitations}
The methodological limitations in this study are due to the following assumptions:
\begin{itemize}
    \item Operational costs: It is assumed that the WF does not incur any operational costs, including start-up costs or energy and reserve provision costs.
    \item WT availabilities: The availability of WTs is not considered in the optimisation problem formulation; however, the unavailability of certain WTs in the WF would hinder its power production. 
This aspect could be incorporated by considering the Mean Time to Failure (MTTF) and Mean Time to Repair (MTTR), together with the WF's accessibility, quantified through the Mean Time to Wait (MTTW) for a suitable weather window \cite{centeno-telleria_impact_2024}, to generate WT availability scenarios. 
    \item Risk aversion: In this study, the WPP is assumed to be risk-neutral. 
A risk management strategy, such as Conditional Value-at-Risk (CVaR), could be included to reduce the risk of having extremely high imbalance losses by specifying a desirable weighting between expected profit and risk.
    \item Prices and penalties are known a priori: The WPP is assumed to act as a price taker in the electricity market, with the associated implications discussed above. 
In reality, providers are generally price makers, and wind power can be subsidised (as is the case for most of it in GB \cite{edmunds_participation_2019}), which makes prices for it different.
In addition, market prices are also an uncertain parameter, which can be forecasted using recent market observations \cite{botterud_risk_2010} or modelled as a log-normal distribution for each hour \cite{shafie-khah_optimal_2014}, providing a basis for generating price scenarios.
An envisaged application of the proposed method is the determination of an optimal offering curve for a WPP participating in the DA market \cite{lee_bivariate_2018}. 
In this process, a price is fixed (for a given hour), and the optimisation problem finds the corresponding optimal energy (quantity) to be offered at that price, forming a price-quantity pair. 
By repeating this procedure over a range of price levels, a set of optimal price-quantity pairs is obtained, which collectively constitute the offering-curve submitted to the market. 
\item Wake modelling: This study employs steady-state engineering wake models.
While these models provide computational efficiency, they do not capture the transient dynamics of atmospheric conditions and turbine operations. 
A more accurate representation of wake effects could be achieved using dynamic wake models, such as the one presented in \cite{becker_dynamic_2025}.
\item Suitability of wake steering: The assessment in this study is based solely on the energy gains from power production.
To obtain a more comprehensive understanding of the impact of wake steering on the levelised cost of energy for offshore wind projects, structural load analysis and its impact on turbine lifetime must also be incorporated.

\end{itemize}

\subsection{Key takeaways and Result implications} 
\label{takeaways}
The main findings from this study are the following:
\begin{itemize}
    \item The Power Curve method yields power production and expected income estimates 12-13\% higher than those of the Baseline method, resulting in 3\% lower actual revenue due to imbalance penalties, despite the Baseline method submitting lower bids. 
    \item Employing wake steering (WRC approaches) yields a 1-2\% gain in income relative to the Baseline method, highlighting the economic potential of wake steering in short-term electricity markets.
      \item Although both WRC strategies yield identical results in this case study, the underlying reasons for this outcome are analysed to better understand the conditions under which they might produce different results.
\end{itemize}
The findings of this study demonstrate that the proposed methodology reliably accounts for wake effects while optimising bidding and trading of wind power in short-term electricity markets. 
Furthermore, the computational time required for the optimisation process is compatible with DA market planning timelines.
For instance, the GB hourly DA market consists of 24 tradable contracts, one for each delivery hour of the following day, traded individually or in hourly blocks, with the order book closing at 09:20 GMT and results published from 09:30 GMT \cite{noauthor_epex_2024-1}.
In applying the proposed method to determine an optimal bidding strategy, a WF operator must prepare and submit a bid curve (price-quantity pairs per hour) within a single working day.
As the computation for each pair is independent and can be parallelized, the total computational time is compatible with the required time-frame. 
Results suggest that current models inaccurately estimate wind power production when participating in short-term electricity markets.
When estimating wind power production using a power curve, higher estimations can result in substantial penalties for imbalances.
However, available production can be underestimated if the WF lacks the incorporation of a wake management technique (which is usually the case). 
Employing wake steering yields increases in income as more power is being harvested from the WF.
Results also align with the demonstrated sensitivity of wind power production to wind direction, underscoring the importance of this parameter and further highlighting the utility of wake steering.
The findings of this study is relevant in the context of wind power bidding and trading in short-term electricity markets, introducing a new perspective by reporting the impacts of wake effects in WFs in this field.

\section{Conclusion}
\label{conclusion}
In the current context of energy and environmental crisis, public policies and incentives are promoting new wind power installations around the world. 
As wind energy penetration increases, WPPs are increasingly being asked to perform ancillary grid services, that are required by TSOs to maintain integrity, stability, and power quality of the power system. 
Simultaneously,  the presence of wakes effects in WFs pose challenges in terms of power losses and increased fatigue loads on WTs. 
Energy and reserve services provided by WPPs are traded in electricity markets, and accurate estimation of wind power production is crucial for effective participation in short-term electricity markets.
Existing models represent an upper limit of generation using the power curve provided by the turbine manufacturer neglecting wake effects, which can lead to high imbalance costs in the balancing market.
This paper presents a WF energy and reserve scheduling model that incorporates wake steering to optimise participation in DA energy and reserve markets, applied to the GB power system.
A market framework is introduced to incentivise WPPs to maximise profits through coordinated bidding in both markets.
The proposed methodology employs a two-stage stochastic programming approach where the first stage determines market bids and the second stage evaluates operational decisions under uncertainty.
Scenarios for wind speed, wind direction, and FRR demand are generated to capture the variability inherent in wind and system conditions. 
The effectiveness of the approach is evaluated through a case study based on a simulated version of the London Array offshore WF. 
Since direct measurements at the site were unavailable, atmospheric conditions were modelled using wind data from a nearby location with comparable meteorological characteristics, ensuring realistic wind profiles for simulation purposes. 
The turbine layout and technical specifications closely match those of the actual installation, providing high fidelity for evaluating wake interactions and control strategies. 
Numerical results demonstrate the impact of wake effects on bidding and trading strategies, providing a novel perspective on wind power market participation.
\subsection{Future work} 
\label{future}
Future work will aim to improve the overall problem formulation by addressing the limitations outlined in Section \ref{limitations}.
In particular, the current model does not consider the additional structural loads and mechanical wear on turbines induced by wake steering. 
Subsequent studies should therefore include a qualitative assessment and estimation of the associated fatigue and cost impacts on long-term economics.
Future research directions also include the development of dynamic yaw control strategies for real-time ancillary service provision, along with a more comprehensive analysis of their effects on turbine fatigue and WF lifespan under varying operational conditions. 

\section{Acknowledgments}
\label{acknowledgments}
The authors would like to thank the five anonymous reviewers for their valuable comments and constructive suggestions, which have helped to improve the quality and clarity of this paper. 

\newpage 
\appendix
\section{Sensitivity analysis}
\label{app1}
\begin{figure}[h] 
\centering
\includegraphics[width=\textwidth,keepaspectratio]{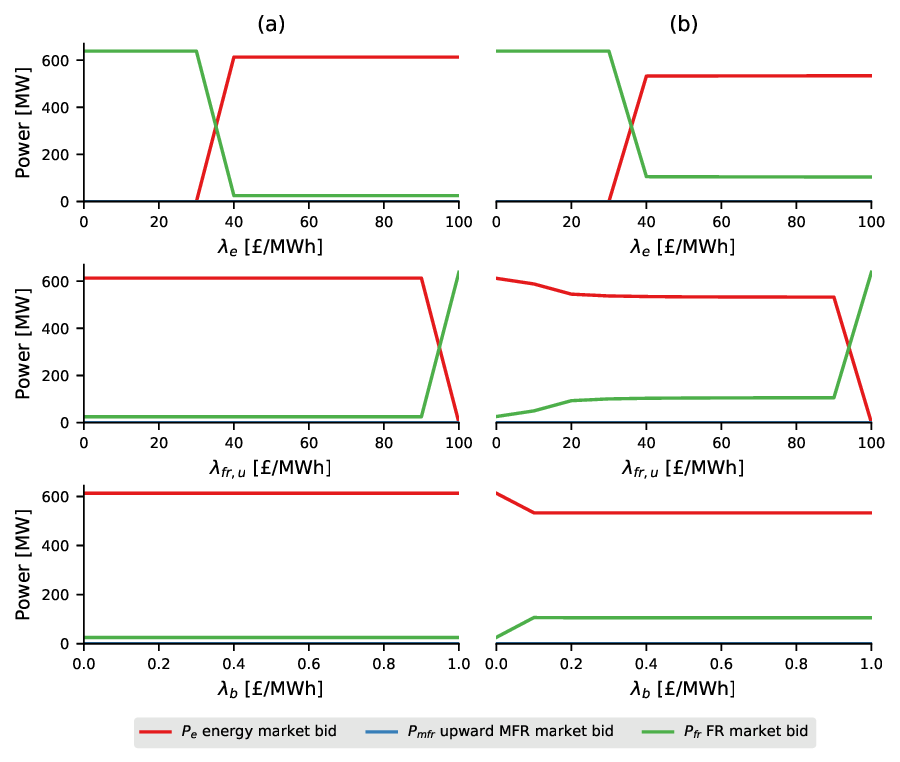}
\caption{Variation of energy and reserve market offers under varying energy prices, FR utilisation payments, and imbalance penalties: (a) stable wind conditions (wind speed: $\mu$ = 14.3 \si[per-mode=symbol]{\meter\per\second}, $\sigma$ = 0.8 \si[per-mode=symbol]{\meter\per\second}; wind direction: $\mu$ = \ang{-113.8}, $\sigma$ = \ang{3.3}); and (b) high wind variability (wind speed: $\mu$ = 12.4 \si[per-mode=symbol]{\meter\per\second}, $\sigma$ = 3.5 \si[per-mode=symbol]{\meter\per\second}; wind direction: $\mu$ = \ang{62.5}, $\sigma$ = \ang{80.5}).}\label{fig:a1}
\end{figure}

Fig. \ref{fig:a1} presents the results of the sensitivity analysis conducted to assess the impact of the variation of energy prices, FR utilisation payments, and imbalance penalties under different cases of wind conditions.

\end{document}